\documentclass[preprint]{article}
\usepackage{graphicx}

\newcommand{\be}{\begin{equation}}
\newcommand{\ee}{\end{equation}}
\newcommand{\ba}{\begin{eqnarray}}
\newcommand{\ea}{\end{eqnarray}}

\begin{document}
\setlength{\textheight}{8.5 in}
\begin{titlepage}

\vskip 5mm

\vskip 1mm
\vskip 20mm

\begin{center}
{\Large\bf { Conformal Bootstrap Analysis for Single and Branched Polymers  }}
\vskip 6mm
 S. Hikami 

\vskip 5mm

Mathematical and Theoretical Physics Unit,
Okinawa Institute of Science and Technology Graduate University,
Okinawa, Onna 904-0495, Japan; hikami@oist.jp

\end{center}
\vskip 20mm
\begin{center}
{\bf Abstract}
\end{center}
\vskip 3mm
The  determinant method  in the conformal bootstrap is applied for  the critical phenomena of a single polymer in arbitrary $D$ dimensions. The scale dimensions (critical exponents) of the polymer ($2< D \le 4$) and the branched polymer ($3 < D \le 8$)  are 
obtained from the small determinants.  It is known that the dimensional reduction of  the branched polymer in $D$ dimensions to Yang-Lee edge
singularity in $D$-$2$ dimensions  holds exactly.  We examine this equivalence by  the small determinant method.
 \end{titlepage}

\newpage
\section{Introduction}
\vskip 2mm
The conformal field theory in arbitrary dimensions was developed long time ago \cite{Ferrara1973,Polyakov1974}, and the modern numerical approach was initiated by \cite{Rattazzi2008}. 
The studies by this conformal bootstrap method led to  promissing results  for various symmetries in general dimensions $D$.
The  review article \cite{Poland2018} includes conformal bootstrap developments where recent references  may be found.
   
    Instead of taking many relevant operators,   the determinant method with small prime operators provides interesting results for the non-unitary cases. The determinant method is applied on  Yang-Lee edge singularity  with considerable accuracies \cite{gliozzi2013,gliozzi2014,Hikami2017}. The polymer case is known as another non-unitary case. The method of
    finding a kink at the boundary of the unitary condition for O(N) vector model \cite{Kos2015,Kos2016} breaks down for $N<1$, and one needs   higher operators for the polymer case, which corresponds to $N=0$ \cite{Shimada2016}.

This paper deals with two different polymers with the determinant method : the single polymer  and the branched polymers in a solvent. They have different upper critical dimensions, 
4 and 8, respectively.
It is well known that the  polymer in a solvent are equivalent to self avoiding walk, which was studied by the renormalization group $\epsilon$ expansion   ($\epsilon = 4 - D$) for $N\to 0$ limit of $O(N)$ vector model \cite{Wilson1972,DeGennes1979}.

 The branched polymer in D dimensions  ($3 < D < 8$) is  equivalent to the Yang-Lee edge singularity in $D-2$ dimensions, as shown  by $\epsilon$ expansion ($\epsilon = 8 -D$)\cite{Lubensky1978,Fisher1978} and by the supersymmetry \cite{Parisi1981}.
 This equivalence  is  further proved exactly  \cite{Brydges2003,Cardy2003}. Due to this rigorous proof, the dimensional reduction $ D \to D-2$  should hold  for $3 < D \le 8$ in the conformal bootstrap analysis. 
 Since Yang-Lee edge singularity for $1< D \le 6$ has been studied  by the conformal bootstrap method \cite{gliozzi2013,gliozzi2014,Hikami2017}, it is interesting to apply the determinant method on the  branched polymer concerning with  the verification of the  equivalence.
 
 We concern with two issues on polymers :
 (i) the critical phenomena of polymers  belong to the logarithmic conformal field theory since the central charge $C$ becomes zero \cite{Gurarie2004,Cardy2013,LeClair2018}, (ii) the method of the replica limit $N\to 0$ is equaivalent to  the use of the supersymmetry \cite{Parisi1981}. The validity of the supersymmetric arguments   has been discussed in a long time for the   the random magnetic field Ising model (RFIM) \cite{Parisi1979}. In RFIM, the dimensional reduction to $D-2$ dimensional pure Ising model will break down at some lower critical dimensions, which has been shown rigorously \cite{Imbrie1984}. Then the lower critical dimension is suggested  to be around three dimensions, above which the supersymmetry argument may be valid \cite{Hikami2018}.   The study of the branched polymer is theoretically interesting from the point of the validity of the supersymmetry.  
The conformal bootstrap method may give a clue  to the relation between the supersymmetry and replica limit.

 In this paper, we evaluate the scale dimensions of the single  polymer and a branched polymer by the determinant method with a small numbers  of the operators.
This study is an extension of a previous analysis of Yang-Lee edge singularity \cite{gliozzi2013,Hikami2017}, in which we have a constraint $\Delta_\phi=\Delta_\epsilon$ due to the
equation of motion in $\phi^3$ theory.  We define the scale dimension of the energy as $\Delta_\epsilon= \Delta_{\phi^2}$, where $\phi$ is the order parameter of $\phi^3$ theory. For the polymers, we have $\phi^4$ theory by the symmetry.   Instead of $\Delta_\phi= \Delta_\epsilon$,  we have an important constraint of the
crossover exponent $\hat \varphi$ \cite{Hikami1974} of O(N) vector model. It has a relation to   $ \Delta_\epsilon$, as
$\Delta_T = \Delta_\epsilon$, where $\Delta_T = D - \hat \varphi/\nu$ ($\nu$ is the critical exponent of the correlation length). The scale dimension of the energy is defined generally by $\Delta_\epsilon = D - \frac{1}{\nu}$. Therefore, for polymers,  we have the crossover exponent $\hat \varphi = 1$, which leads to $\Delta_T = \Delta_\epsilon$.

Although we use these constraints in the determinant method, we extend the analysis  by the introducing small  difference between $\Delta_T$ and $\Delta_\epsilon$ ($\Delta_T\ne \Delta_\epsilon$), which is analogous to "resolution of singularity" by "blow up", to locate the values of the scale dimensions \cite{Hikami2017}.

The bootstrap method uses the crossing symmetry of the four point amplitude.
The  four point correlation function for the scalar field $\phi(x)$ is given by
\be
<\phi(x_1)\phi(x_2)\phi(x_3)\phi(x_4)> = \frac{g(u,v)}{|x_{12}|^{2\Delta_\phi}|x_{34}|^{2\Delta_\phi}}
\ee
and the amplitude  $g(u,v)$ is expanded as the sum of conformal blocks $G_{\Delta,L}$ ($L$ is a spin),
\be
g(u,v) = 1 + \sum_{\Delta,L} p_{\Delta,L} G_{\Delta,L}(u,v)
\ee
The crossing symmetry of $x_1\leftrightarrow x_3$ implies 
\be\label{crossing}
\sum_{\Delta,L} p_{\Delta,L} \frac{v^{\Delta_\phi}G_{\Delta,L}(u,v)- u^{\Delta_\phi}G_{\Delta,L}(v,u)}{u^{\Delta_\phi}-v^{\Delta_\phi}} = 1.
\ee

In the previous paper \cite{Shimada2016}, a polymer case has been studied from the kink behavior at the unitary boundary.
Minor method is consist of the derivatives at the symmetric point $z=\bar z= 1/2$ of (\ref{crossing}). 
By the change of variables $z=(a+ \sqrt{b})/2$, $\bar z= (a-\sqrt{b})/2$, derivatives  are taken about $a$ and $b$. 
Since the numbers of equations  become larger than the numbers of the truncated variables $\Delta$,
we need to consider the minors for the determination of the values of $\Delta$. The matrix elements of minors are expressed by,
\be\label{bootstrap}
f_{\Delta,L}^{(m,n)}= (\partial_a^m \partial_b^n \frac{v^{\Delta_\phi}G_{\Delta,L}(u,v)- u^{\Delta_\phi}G_{\Delta,L}(v,u)}{u^{\Delta_\phi}-v^{\Delta_\phi}})|_{a=1,b=0}
\ee
and the minors of $2\times 2$, $3\times 3$ for instance, $d_{ij}$, $d_{ijk}$ are  the determinants such as
\be\label{dijk}
d_{ij}= {\rm det}(f_{\Delta,L}^{(m,n)}),\hskip 3mm
d_{ijk} = {\rm det} ( f_{\Delta,L}^{(m,n)} )
\ee
where $i,j,k$ are numbers chosen differently from  (1,...,6), following the dictionary correspondence to $(m,n)$ as  $1\to (2,0)$, $2\to (4,0)$, $3\to (0,1)$, $4\to (0,2)$, $5\to (2,1)$ and $6\to (6,0)$. We use the same notations for the conformal block and for the minors as \cite{Hikami2017}.

\vskip 3mm
\section{Replica limit $N\to 0$ for polymer}
\vskip 2mm
There are many examples of the critical phenomena which are non-unitary. The negative value of the coefficients of the operator product expansion (OPE) leads to non-unitary case.
 For instance, this can be seen in the case of Yang-Lee edge singularity and in the polymers. For such non-unitary critical phenomena, the unitarity condition does not hold, and the direct use
of the application of the unitarity boundary condition does not work. For instance,  O(N) vector model shows a kink behavior of the unitary bound for $N > 1$ \cite{Kos2016}, but it looses the kink behavior for $N < 1$, i.e. a kink becomes a smooth curve. One needs other conditions to determine the anomalous dimensions for polymers, which are realized in the replica limit $N\to 0$ \cite{Shimada2016}.

On the other hand, the determinant method works for the non-unitary Yang-Lee edge singularity \cite{gliozzi2013, Hikami2017}. Therefore, it is meaningful to apply the determinant method to polymers. The aim of this paper is the application of the determinant method for polymers.

However, one needs to solve some difficulties in the replica limit $N\to 0$ for this application. The first one is related to the symmetric tensor operator in (\ref{tensor}), which anomalous dimension (\ref{phi}) is related to the
crossover exponent of O(N) vector model \cite{Hikami1974}.
For O(N) vector model, the four point function is  expressed as \cite{Polyakov1974}
\be
<\phi_i(x_1)\phi_j(x_2)\phi_k(x_3)\phi_l(x_4)> \frac{g(u,v)}{|x_{12}|^{2\Delta_\phi} |x_{34}|^{2\Delta_\phi}}
\ee
with
\ba
g(u,v) &=& 1 + \sum_S \delta_{ij}\delta_{kl} p_{\Delta,L} G_{\Delta,L}(u,v)\nonumber\\
&+& \sum_T(\delta_{il}\delta_{jk} + \delta_{ik}\delta_{jl} - \frac{2}{N} \delta_{ij}\delta_{kl}) p_{\Delta,L}G_{\Delta,L}(u,v)\nonumber\\
&+& \sum_A (\delta_{il}\delta_{jk}-\delta_{ik}\delta_{jl})p_{\Delta,L}G_{\Delta,L}(u,v)
\ea
where $S$ means the singlet sector, $T$ is a tensor sector, and $A$ is an asymmetric tensor sector.
\ba\label{Tpole}
&&S: \hskip 3mm \epsilon(x) = \sum_{a=1}^N :{\phi_a}^2(x):\nonumber\\
&&T: \hskip 3mm \varphi_{ab}(x) = :\phi_a(x) \phi_b(x): - \frac{\delta_{ab}}{N} \sum_{c=1}^N : {\phi_c}^2(x) :
\ea
where $S$ means the singlet sector, $T$ is a tensor sector, and $A$ is an asymmetric tensor sector.
\ba\label{tensor}
&&S: \hskip 3mm \epsilon(x) = \sum_{a=1}^N :{\phi_a}^2(x):\nonumber\\
&&T: \hskip 3mm \varphi_{ab}(x) = :\phi_a(x) \phi_b(x): - \frac{\delta_{ab}}{N} \sum_{c=1}^N : {\phi_c}^2(x) :
\ea

The catastrophic divergence is the factor $1/N$ in the tensor sector $T$ in (\ref{tensor}) in the limit of $N\to 0$.
It is known that the crossover exponent $\phi$ of O(N) vector model becomes one in the replica limit. Therefore, the anomalous dimension of this
symmetric tensor operator $T$, denoted as $\Delta_T$ becomes degenerate to $\Delta_{\phi^2}=\Delta_\epsilon$, since
we have
\be\label{phi}
\Delta_T = D - \frac{\hat \varphi}{\nu}
\ee
where $\hat\varphi$ is a crossover exponent and $\nu$ is a critical exponent for the correlation length.
We have by the definition,
\be
\Delta_\epsilon = D - \frac{1}{\nu},
\ee
and we have following degeneracy, due to $\hat \varphi=1$,
\be
\Delta_T = \Delta_\epsilon
\ee
for the polymer case.
This degeneracy of $\Delta_\epsilon=\Delta_T$ may solve the catastrophe of the replica limit as indicated in \cite{Hogervorst2017}. 

The second catastrophe, which is related to the central charge $C$ gives the logarithmic CFT \cite{Gurarie2004,Cardy2013}. The polymer has central charge $C$=0.  The minor method gives the values of the anomalous dimensions without knowing the OPE coefficients as the equation in (\ref{bootstrap})
indicates.
The central charge $C$ is expressed by  the OPE coefficient as 
\be
C = \frac{({\Delta_\phi})^2}{p_{[D,2]}}
\ee
where $\Delta_\phi$ is an anomalous dimension. $p_{[D,2]}$ is the square of the OPE coefficient of energy momentum tensor, and it has a simple pole when
$C=0$.
It is known in two dimensions  that  the vanishing central charge $C=0$ leads to the logarithmic CFT \cite{Gurarie2004,Cardy2013}. For general dimensions, the situation may be  same but the precise forms of OPE coefficients are unknown.

\section{ Determinant method for a single polymer}
\vskip 2mm
{\bf (i) D=4}
\vskip 2mm
We consider the case of self-avoiding walk or polymer in a solvent. For this case, which corresponds to $N=0$ limit of O(N) vector model, the degeneracy of $\Delta_\epsilon$ and $\Delta_T$ occurs, since
the crossover exponent $\hat\varphi$ becomes exactly one by $\epsilon$ expansion in all orders for $N=0$.
\be
\Delta_\epsilon = \Delta_T.
\ee

At the upper critical dimension $D=4$, we have free field value of $\Delta_\phi=1.0$.  The intersection of loci $d_{123}$ and $d_{124}$ to $\Delta_\epsilon = \Delta_t$ line is shown in Fig.1, in which the polymer's scale dimension  becomes $\Delta_\epsilon = \Delta_T= 2.0$. In Fig.1, on the straight line of $\Delta_\epsilon = \Delta_T$, any point satisfies the condition,
and the value of $\Delta_\epsilon = \Delta_T$ is not determined uniquely. We use the "blow up" technique for this degeneracy by introducing the small parameter, which indicates $\Delta_\epsilon\ne \Delta_T$. The blow up technique is known in the theory of resolution of the singularities \cite{Hironaka}. Then, the intersection of lines  will provide the value of $\Delta_\epsilon = \Delta_T$ at the intersection point. We call this procedure as "blow up".
 In Fig.1, the zero loci of the minors $d_{123},d_{124},d_{134}, d_{234}$ intersect with a straight line of $\Delta_\epsilon=\Delta_T$ at $\Delta_\epsilon =2$. The notation of $3\times 3$ minors $d_{ijk}$ is given by (\ref{dijk}). 
The $3\times 3$ minor, for instance $d_{123}$, is
\be
d_{123}= {\rm det} \biggl(\begin{array}{ccc} f_{\Delta_\epsilon,L=0}^{(2,0)} & f_{(D,2)}^{(2,0)}  & f_{\Delta_T,L=0}^{(2,0)}\\
f_{\Delta_\epsilon,L=0}^{(4,0)} & f_{(D,2)}^{(4,0)} & f_{\Delta_T,L=0}^{(4,0)}\\
f_{\Delta_\epsilon,L=0}^{(0,1)} & f_{(D,2)}^{(0,1)}  & f_{\Delta_T,L=0}^{(0,1)}\end{array} \biggr)
\ee
where $f_{\Delta,L}^{(m,n)}$ is given by (\ref{bootstrap}). We here consider only $N=0$ case.

\begin{figure}
\centerline{\includegraphics[width=0.5\textwidth]{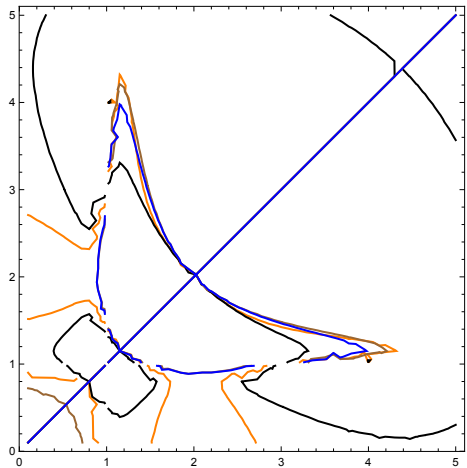}}
\caption{ D=4:  Scale dimensions of polymer. The zero loci of  the $3\times 3$ minor $d_{123}$ (red) , $d_{124}$(brown), $d_{134}$(black) and $d_{234}$(blue) intersect in $D=4$, for   $\Delta_\phi=1.0$ at the point of $\Delta_T=\Delta_\epsilon = 2.0$.  The axis is  (x,y) = $(\Delta_\epsilon, \Delta_T)$. This figure shows the blow up of the degeneracy of $\Delta_\epsilon= \Delta_T$.}
\end{figure}

\vskip 2mm
{\bf (ii) D=3}
\vskip 2mm
For three dimensions, previous conformal bootstrap method gives the values of $\Delta_T=\Delta_\epsilon = 1.2984$, and $\Delta_\phi = 0.5141$ \cite{Shimada2016}, and Monte Carlo gives
$\Delta_T = 1.2982, \Delta_\phi=0.5125$ \cite{Clisby2010}. The $\epsilon$ expansion gives the estimation as $\Delta_T= 1.2999$ and $\Delta_\phi = 0.5142$ \cite{Guida1998}.
The bootstrap method  has an estimate of $\Delta_T$ \cite{Shimada2016} which is close to the result of $\epsilon$ expansion .

In $D=3$, if we adapt the value of $\Delta_\phi=0.514$, which is taken from \cite{Shimada2016}, the intersection of $d_{123}$(orange), $d_{124}$(brown),$d_{134}$(black) and $d_{234}$(blue) are shown in Fig.2, in which $\Delta_\epsilon = \Delta_T= 1.3$ is obtained from $d_{124}$ (brown line).
\begin{figure}
\centerline{\includegraphics[width=0.5\textwidth]{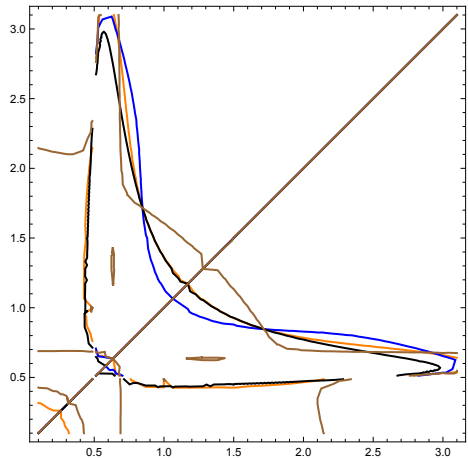}}
\caption{ D=3: Scale dimensions of polymer. The zero loci of  the $3\times 3$ minor $d_{123}$ (orange) , $d_{124}$(brown), $d_{134}$(black) and $d_{234}$(blue).   The point $\Delta_T = \Delta_\epsilon = 1.3$ is realized in $d_{124}$(brown). The axis is  (x,y) = $(\Delta_\epsilon, \Delta_T)$ .}
\end{figure}
 
\vskip 2mm
\noindent{\bf{Table 1. Scale dimensions of a single polymer}}. 
\vskip 1mm
The value of $\Delta_T$ is obtained from 
the zero loci of $3\times 3$ minors.  For D=2, exact values are $\Delta_\phi=5/48, \Delta_\epsilon = 2/3$.\\
\begin{tabular}[t]{|c|c|c|c|}
\hline
 $D $& $\Delta_\phi$ & $\Delta_T=\Delta_\epsilon$ &$ \Delta_\epsilon$($\epsilon$ expansion, exact value)\\
 \hline
2& 0.1  &  0.7& 0.666 \\
3& 0.514 & 1.3 & 1.299\\
3.5 & 0.75 &  1.57 & 1.57\\
4 & 1.0  &  2.0& 2\\
\hline
\end{tabular}
\vskip 2mm

\vskip 2mm

The $\epsilon$ expansion of $\Delta_\epsilon$  ($\epsilon= 4-D$) for the polymer case, which is obtained by the limit $N \to 0$ in the expression of O(N) vector model, is given by
\cite{Wilson1972}
\be
\Delta_\epsilon = 2 -  \frac{3}{4}\epsilon + \frac{11}{128}\epsilon^2 + O(\epsilon^3)
\ee
which becomes $\Delta_\epsilon = 1.57$ for $D=3.5 (\epsilon= 0.5)$. For D=3, $\epsilon $ expansion by Borel Pad\'e analysis gives $\Delta_\epsilon = 1.3$ \cite{Guida1998}, which are close to the values obtained by $3\times 3$ determinant  $d_{124}$ in Fig.2. There are splits (or jump) of the intersection points along the diagonal line around $\Delta_\epsilon = 1.3$. Such split behavior
has been observed also in the analysis of Yang-Lee edge singularity at the critical dimension $D_c$, where $\Delta_\phi =0$ and the central charge $C$ is estimated as $C=0$.
We took the maximum value of $\Delta_T$ of the splitting points in Fig.2. The Yang-Lee edge singularity is indicating  a reasonable value $D_c$ by taking the maximum point of the splitting for blow up \cite{Hikami2017}.
In polymer case, also the maximum of the split values in $D=3$ is close to the $\epsilon$ analysis (Table.1.). Further investigation of this split (or jump) behavior is required by the systematic analysis,  and it remains as a future work. The determinant method using a small rank matrix gives a rough estimation, and the result depends on the
choice of $d_{ijk}$. Recent article \cite{LeClair2018} also pointed out that special choice of the determinant is better for the estimation of the anomalous dimension $\Delta_\epsilon$.
The method for the estimation of the error bar was suggested in \cite{Li2017}.

The central charge $C=0$ suggests the pole of the OPE coefficient of the energy momentum tensor, which leads to the logarithmic CFT behavior.
The pole for $N\to 0$ in (\ref{Tpole}) is related to the degeneracy of $\Delta_\epsilon = \Delta_T$ as explained in \cite{Hogervorst2017} for polymer case;
the OPE coefficient of $\Delta_T$  has a pole.

\vskip 2mm
\newpage
\section{Branched polymer}
\vskip 2mm
There is a remarkable  equivalence, so called as dimensional reduction, between a branched polymer in D dimensions ($3 < D < 8$) and Yang-Lee edge singularity in D-2 dimensions ($1 < D < 6$); the critical exponents become same.  The branched polymer is described by $\phi^3$ theory, but the upper critical dimensions is known to be 
 8  due to the disorder. The $\epsilon=8-D$ expansion for the critical exponent agrees with the exponent of Yang-Lee edge singularity in $\epsilon = 6-D$ expansion , for instance the critical exponent $\nu$ becomes same for both models \cite{Lubensky1978,Fisher1978}.

The action of the branched polymer has the branching terms in addition to the self-avoiding term (single polymer). We write this action for the $p$-th branched polymer as $N$-replica field theory \cite{Cardy2013}
\be
S = \int d^D x ( \frac{1}{2}\sum_{\alpha=1}^N [ (\nabla \phi_\alpha)^2 - \sum_{p=1}^\infty u_p \phi_\alpha^p] + \lambda(\sum_{\alpha=1}^N \phi_\alpha^2)^2 )
\ee
The term $\phi_\alpha^p$ represents the $p$-th branched polymer. After the rescaling and by neglecting irrelevant terms, the following action is obtained

\be
S= \int d^D x (\frac{1}{2}\sum_{\alpha=1}^N [(\nabla\phi_\alpha)^2 + V(\phi_\alpha)] + g \sum_{\alpha,\beta=1}^N \phi_\alpha \phi_\beta)
\ee
where $V(\phi)$ = $t \phi - \frac{1}{3}\phi^3 + O(\phi^4)$.

As same as before, $N=0$ replica limit of O(N) vector model is applied  for this branched polymer.  The condition of $\Delta_\epsilon=\Delta_T$ is also essential  in a branched polymer problem. We find the several fixed points in the blow up plane of $\Delta_\epsilon,\Delta_T$.
In the branched polymer case, we find new triple degeneracy
\be
\Delta_\epsilon=\Delta_T = \Delta_\phi+1.
\ee
The last term of 1 is a trivial term due to the definition of $\Delta_\phi$ in D dimensions. The $\epsilon$ expansion of the branched polymer becomes \cite{Lubensky1978,Fisher1978},

\be
\eta = - \frac{1}{9}\epsilon
\ee
where $\epsilon= 8 - D$.
The scaling dimension $\Delta_\phi$ is defined by
\be
\Delta_\phi ({\rm branched \hskip 1mm polymer}) = \frac{D - 2 + \eta}{2} 
\ee
In above formula, if we put $D \to D-2$, and put $ \epsilon = 6 -D$, then we get
\be
\Delta_\phi ({\rm Yang}- {\rm Lee \hskip 1mm edge \hskip 1mm singularity}) = 2 - \frac{5}{9}\epsilon
\ee
where $\epsilon = 6-D$.
This shows exactly the dimensional reduction relation between a branched polymer and Yang-Lee edge singularity.

The exponent $\nu$ of Yang-Lee edge singularity $(\epsilon= 6 - D$) is
\be
\frac{1}{\nu}= \frac{1}{2}(D+ 2 - \eta) = \frac{1}{2}(8 - \epsilon + \frac{1}{9}\epsilon) = 4 - \frac{4}{9}\epsilon
\ee
This leads in Yang-Lee edge singularity,
\be
\Delta_\epsilon = D - \frac{1}{\nu} = (6-\epsilon) - (4 -\frac{4}{9}\epsilon) = 2- \frac{5}{9}\epsilon = \Delta_\phi
\ee
The condition $\Delta_\epsilon=\Delta_\phi$ is a necessary condition for Yang-Lee edge singularity due to the equation of motion.

By the dimensional reduction, the values of exponents $\eta$ and $\nu$ of branched polymer become same as Yang-Lee edge singularity.
The scale dimensions of $\Delta_\epsilon$ and $\Delta_\phi$, however, become different since they involve the space dimension $D$ explicitly.
In a branched polymer of $D=8$,
\be\label{equivalence}
\Delta_\epsilon =4, \hskip 3mm
\Delta_\phi = 3
\ee
where
for Yang-Lee edge singularity of D=6,
\be
\Delta_\epsilon= 2,\hskip 3mm \Delta_\phi = 2.
\ee

In general dimension $D \le 8$, from the equivalence to Yang-Lee edge singularity, we have
\be\label{super}
\Delta_\epsilon = \Delta_\phi + 1
\ee
as shown in (\ref{equivalence}) for $D=8$. This relation is related to the supersymmetry as discussed in \cite{Shimada2016,Bashkirov2013,Fei2016}.

\vskip 2mm

  We get the following relations
 \ba\label{brYL1}
&& \Delta_\phi ({\rm branched\hskip 1mm polymer \hskip 1mm in \hskip 1mm D \hskip 1mm dim.}) = \Delta_\phi ({\rm Yang\hskip 1mmLee \hskip 1mm in \hskip 1mm D-2 \hskip 1mm dim.}) + 1,\nonumber\\
&& \Delta_\epsilon ({\rm branched\hskip 1mm polymer \hskip 1mm in \hskip 1mm D \hskip 1mm dim.}) = \Delta_\epsilon ({\rm Yang\hskip 1mm Lee \hskip 1mm in \hskip 1mm D-2 \hskip 1mm dim.}) + 2.\nonumber\\
 \ea

In Fig.3, the intersection map of the loci of minors for the branched polymer  is shown. The contour of zero loci $d_{123}(blue)$,$d_{124}(brown)$,$d_{134}(green)$,$d_{234}(red)$ are shown in different colors.  The fixed point of $\Delta_\epsilon = 4$ and $\Delta_\phi=3$ in (\ref{equivalence}) is obtained, which values are consistent with the Yang Lee edge singularity by the dimensional reduction. The parameter
of Q(spin 4) is chosen as 10.
The fig. 3 shows the map of (x,y) = $(\Delta_\phi,\Delta_\epsilon$). There are singular lines in Fig.3. The horizontal line at $\Delta_\phi=6$ is due to the degeneracy of $\Delta_\epsilon = \Delta_T=6$ and other horizontal line at $\Delta_\epsilon =3$ is due to the pole of $\Delta_\epsilon= (D-2)/2$.

We confirm the dimensional reduction to Yang-Lee model in $D-2$ dimension for $4 < D < 8$ by $3\times 3$ determinant method. 

In Fig.4, the branched polymer in D=8 is considered in the blow up map of $(\Delta_\epsilon,\Delta_T)$ with $\Delta_\phi=2$. There is a fixed point at $\Delta_\epsilon = \Delta_T= 4.0$ for the branched polymer. This corresponds to
 Yang-Lee edge singularity  in D=6 ($\Delta_\epsilon = \Delta_\phi=2.0)$ due to the dimensional reduction. These values satisfy ($\ref{brYL1}$).

In Fig.5, the branched polymer in D=7 is shown with $\Delta_\phi=1.4255$. The fixed point can be read as  $\Delta_\epsilon=\Delta_T=3.4. $ This value corresponds to $\Delta_\epsilon = 1.4$ of Yang-Lee edge singularity in D=5.

\begin{figure}
 \centerline{\includegraphics[width=0.5\textwidth]{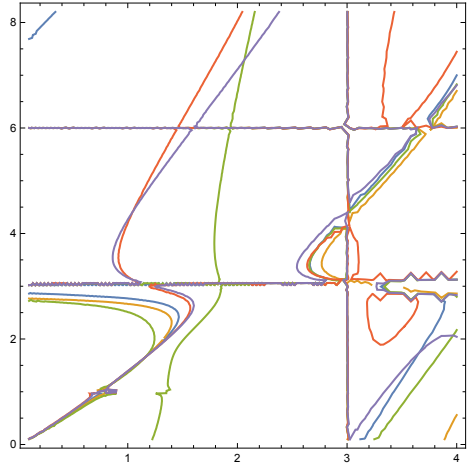}}
\caption{ Branched polymer in D=8 : The fixed point $\Delta_\epsilon$ =4, $\Delta_\phi=3$  is obtained for the branched polymer in (\ref{equivalence}). These values agree with  Yang-Lee edge singularity at D=6  by the dimensional reduction.
 The axis is $(x,y)= (\Delta_\phi,\Delta_\epsilon)$}
\end{figure}
\vskip 2mm

\begin{figure}
\centerline{\includegraphics[width=0.5\textwidth]{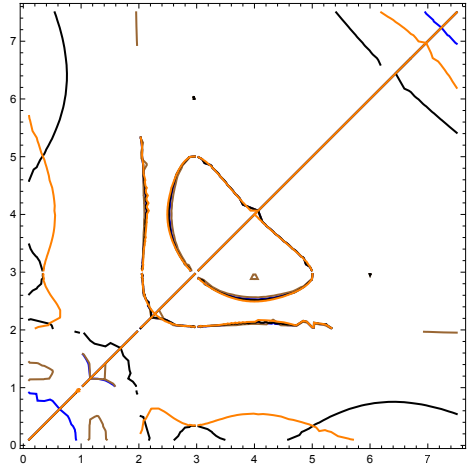}}
\caption{Branched polymer in D=8: The zero loci of  the $3\times 3$ minor .$d_{123}$ (black), $d_{134}$(brown), $d_{124}$(orange), $d_{234}$(blue) in a blow up plane ($\Delta_T\ne\Delta_\epsilon$). The fixed point appears at
$\Delta_\epsilon=\Delta_T$= 4.0. The axis is  (x,y) = $(\Delta_\epsilon, \Delta_T)$ .}
\end{figure}

\begin{figure}
\centerline{\includegraphics[width=0.5\textwidth]{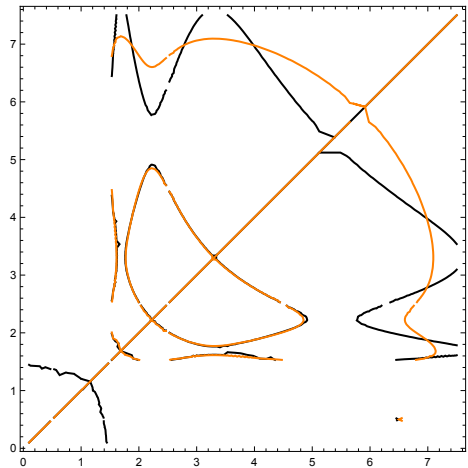}}
\caption{Branched polymer in D=7: The zero loci of  the $3\times 3$ minor $d_{123}$ (black) and $d_{124}$(orange) in a blow up plane. The axis is  (x,y) = $(\Delta_\epsilon, \Delta_T)$. The intersection point in the blow up map is
$\Delta_\epsilon = 3.4$, which is consistent with the dimensional reduction relation between branched polymer and Yang-Lee edge singularity.}
\end{figure}



\vskip 2mm
\newpage

\section{Summary}
\vskip 2mm

We have analyzed  polymers: a single polymer and  a branched polymer, and we
 have found that they are characterized by the degeneracies of the primary operators,
 $\Delta_\epsilon=\Delta_T$, which value is obtained in a blow up plane as an intersection point.
 
 For a single polymer, the scaling dimension $\Delta_\epsilon$ is obtained from the intersection of the zero loci of  $3\times 3$ minors with rather good accuracy (Table.1).

We find in the branched polymer case the exact relation  of $\Delta_\epsilon = \Delta_\phi + 1$ in the determinant method with good numerical accuracy. 
 This relation is consistent with the relation of the dimensional reduction between the branched polymer and Yang-Lee edge singularity. In Yang-Lee edge singularity, by the equation of motion we have $\Delta_\epsilon = \Delta_\phi$. The relation $\Delta_\epsilon = \Delta_\phi + 1$ is a characteristic relation in
 the supersymmetry theory \cite{Bashkirov2013,Shimada2016,Fei2016}, where  Grassmann coordinates give the dimensional reduction (-2) \cite{Parisi1981}.

The validity of the dimensional reduction in random field Ising model (RFIM) has been discussed for a long time, and it is known that the reduction to pure Ising model does not work in the lower dimensions.
We will discuss this problem by the conformal bootstrap determinant method in a separate paper \cite{Hikami2018}.

\vskip 3mm

\vskip 5mm
{\bf Acknowledgements}
\vskip 3mm
Author is thankful to Nando Gliozzi for the discussions of the determinant method.  He  thanks Edouard Br\'ezin
 for the useful discussions of the dimensional reduction problem in branched polymers. This work is supported by JSPS KAKENHI Grant-in-Aid 16K05491. 
\newpage

\newpage

\end{document}